# X-ray phase contrast imaging of biological specimens with tabletop synchrotron radiation

S. Kneip[1,2], C. McGuffey[2], F. Dollar[2], M.S. Bloom[1], V. Chvykov[2], G. Kalintchenko[2], K. Krushelnick[2], A. Maksimchuk[2], S.P.D. Mangles[1], T. Matsuoka[2], Z. Najmudin[1], C.A.J. Palmer[1], J. Schreiber[1], W. Schumaker[2], A.G.R. Thomas[2], V. Yanovsky[2]

[1]*The Blackett Laboratory, Imperial College London, London, SW7 2BZ, UK*
[2]*Center for Ultrafast Optical Science, University of Michigan, Ann Arbor, MI, 48109, USA*

**Since their discovery in 1896, x-rays have had a profound impact on science, medicine and technology. Here we show that the x-rays from a novel tabletop source of bright coherent synchrotron radiation can be applied to phase contrast imaging of biological specimens, yielding superior image quality and avoiding the need for scarce or expensive conventional sources.**

The x-ray tube is still the most common source of x-rays[1], but cutting-edge medical applications demand high quality beams of x-rays, the production of which often requires large, expensive synchrotron x-ray facilities[2]. Our scheme deploys a recently demonstrated tabletop source of 1-100 keV synchrotron radiation[3]. Here, in order to produce high quality x-rays, electrons are accelerated and wiggled analogously to a conventional synchrotron, but on the millimeter rather than tens of meter scale. We use the scheme to record absorption and phase contrast images of a tetra fish, damselfly and yellow jacket, in particular highlighting the contrast enhancement achievable with the simple propagation technique of phase contrast imaging.

X-rays have a much shorter wavelength than visible light, can penetrate matter and image the interior of solid objects. Image contrast is obtained through preferential absorption of x-rays in dense regions of the sample (absorption contrast)[1], or through bending of the wavefront in regions of the sample with differing refractive index decrement (phase contrast)[4]. For medical radiography, an absorption contrast contact image is recorded of the specimen, which is in contact with the detector. High contrast is limited to dense tissue such as bones. Image contrast from low-density tissue can be enhanced using radioopaque and radiolucent agents, but administration is invasive and/or limited to certain applications[5]. Phase sensitive radiography sidesteps the need for radioopaque agents to visualize soft tissue, and has the additional benefit that the dose can be reduced by using harder, more transmissive x-rays[6].

The huge potential of phase contrast imaging for medical diagnostics has been realized for quite some time, documented by numerous animal studies[7-9]. Progress has been held back due to the lack of suitable x-ray sources with the necessary spatial coherence properties and/or due to the requirement for cumbersome imaging techniques. Interferometric and refractive techniques require optics, are limited in their field of view, e.g. by requiring monolithic crystals, and/or require scanning data techniques[6, 10, 11]. Spatial coherence $L_{trans}$ increases proportionally with the distance from the source $u$ and with the inverse of the source size $w$. Conventional x-ray tubes must therefore be apertured, increasing the required exposure time. Development in microfocal x-ray tubes offers improvements, but brightness is limited due to anode material and cooling. With the advent of high power lasers, various schemes have been studied for phase contrast imaging[12-14]. Conventional synchrotrons continue to be the ideal source for phase contrast imaging[2] but access is limited.

As shown in figure 1a, the x-ray source discussed here is based on focusing a pulsed high power laser into a millimeter-sized plume of helium gas, which is immediately ionized and turned into a plasma (see methods). Within the plasma, electrons are accelerated and wiggled analogously to a conventional synchrotron. The observation of an x-ray beam, originating from the interaction and pointed along the laser direction is correlated with the electron beam. Depending on experimental parameters, the x-ray beam divergence is measured to be 5−15 mrad, the 1/e2 x-rays intensity source size is 1 − 3 μm and the spectrum is synchrotron like with average photon energy (critical energy) of $E_{crit} \simeq$ 10-40 keV. Each laser pulse delivers a 30 fs flash of $\simeq 10^6$ photons mrad$^{-2}$. This corresponds to a peak brightness of $10^{22}$ photons per (second mm$^2$ mrad$^2$ 0.1%BW), which is comparable to conventional 3rd generation synchrotrons and makes possible high contrast imaging in a single shot. More details of the experiment are described elsewhere[3].

Due to the small source size, the x-ray beam has an appreciable degree of spatial coherence. This enables phase contrast imaging with a simple propagation technique (see methods). To achieve contrast enhancement with the propagation technique, the image distance (specimen to detector) has to be increased[2].

With figure 1b-d we demonstrate that this x-ray source can produce detailed radiographs of biological specimens with a single 30 fs exposure. Contact images of a tetra fish (figure 2b) and damselfly (figure 2c-d) were taken at a distance of $v$=2.79 m from the source, where the specimen is in contact with the detector at $u$=$v$=2.79 m from the source. Absorption contrast from the absorbing skeleton of the tetra is good but absorption contrast from the non-absorbing thin wings and uniformly absorbing exoskeleton of the damselfly is poor. Using the propagation technique, with the damselfly at $u$=0.44 m and the detector at $v$=1.83 m from the source, the path-integrated phase change of the spatially coherent x-rays through the sample leads to enhanced edge contrast, revealing the fine structure of the wing, the legs and exoskeleton, as shown in figure 1d. The ratio of the cross section for phase shift to absorption is greater than 100 for 18 keV x-rays for Z<15[6]. The transverse coherence length $L_{trans}$ is approximately 10 μm for 10 keV x-rays 0.5 m from a 1 μm source. The figure 3b shows a phase contrast enhanced image of a yellow jacket taken in a single shot 30 fs exposure in the same configuration as figure 1d.

Resolution of the contact radiographs is limited by the 13 μm pixel size of the ccd detector which is much larger than the x-ray source size. Fine details of the skeleton and fins of the fish can be noticed in figure 1b and an intensity lineout across the caudal fin (shaded yellow box in figure 1b) is plotted in figure 2a. The spines and rays of the fin are resolved, demonstrates imaging resolution of at least ≃120 μm.

Due to the increased image distance necessary for propagation phase contrast imaging, the ccd detector is recording a $M$=$v/u$≃4.2 magnified image of the specimen, when compared to the contact images. This has two consequences. Firstly, to capture the entire specimen, multiple sub-images have to be joined, as indicated by the red dashed lines in figure 2. Secondly, the effective detector pixel size 13 μm /4.2=3.1 μm is now on the order of the x-ray source size, permitting a greater image resolution. Thus, to allow for a fair comparison of image quality obtained with absorption and phase technique, we have artificially increased the pixel size of the phase image of figure 1d to 13 μm. Intensity lineouts across the leg of the damselfly (shaded yellow box in figure 1c,d) are plotted in figure 2a. This demonstrates clear edge enhancement and imaging resolution at the few pixel level (≃70 μm) for the phase contrast image 1d. When printing a single shot exposure raw data image with pixel size left unaltered, even greater detail of the specimen, reminiscent of the compound eye, can be seen in figure 2b.

The demonstrated tabletop source of synchrotron radiation may also be suitable for lensless imaging and tomographic reconstruction of 3D phase and absorption information[9]. This would require the broad synchrotron spectrum to be monochromatized. Laser developments proposed in the near-future (e.g. diode pumping) may enable the repetition rate of our system to be increased from 0.1 to 100 Hz. This can compensate any loss of average brightness incurred through monochromatization. The flexibility of a laser-based source combined with a precise narrow x-ray beam would facilitate scanning imaging or multiple views without moving the object. The demonstrated scalability to hard x-rays (>50 keV, ref [3] and therein) complements existing sources and opens the possibility to even do phase contrast imaging of dense objects such as bones[15].

Another unique benefit of this source is the ultrashort pulse duration which can freeze motion blur (the heart beat of a mouse is 10 Hz) and also allow direct study on the timescale of molecular interactions. The absolute time synchronization can enable optical pump-probe experiments. Hence with further development, this tabletop synchrotron source will offer a cheap and compact route to make advanced imaging schemes more commonly available.

**Acknowledgements** This work was partially supported by the US National Science Foundation through the Physics Frontier Center FOCUS Grant No PHY-0114336 and the NSF/DNDO Grant No. 0833499.

**Author Contributions** The experiment and analysis was carried out in main by SK, CM and FD with contributions from MSB, SPDM, TM, CAJP and WS. VY, GK and VC operated the laser, SK, CMG, KK, ZN, JS and AGRT contributed to planning, interpretation and manuscript preparation.
**Competing Interests** The authors declare that they have no competing financial interests.
**Correspondence** Correspondence should be addressed to SK (email: stefan.kneip@imperial.ac.uk).

**Figure 1:** Schematic of the experimental setup and results. (a) A high power laser is focused into a tenuous gas jet, creating a miniature plasma accelerator and wiggler, analogous to the conventional accelerator and wiggler of a synchrotron x-ray light source. The emerging electron and x-ray beams are separated with a magnet. (b-d) The x-ray beam can be used to image biomedical specimens. X-ray contact radiograph (absorption contrast) yields good contrast for the case of a tetra fish (b), du to its highly absorbing skeleton, but poor contrast of a damselfly (c), due to poorly absorbing wings (d) Exploiting the coherent quality of our tabletop synchrotron x-rays, the image quality of the damselfly can be improved by propagation the x-rays 1.4 m from the specimen to the detector (phase contrast). The phase contrast images are taken in a single shot 30 fs exposure. Yellow boxes are explained in the text.

**Figure 2:** Improved image quality with phase contrast imaging. (a) Single shot 30 fs exposure of the head of the damselfly. The details of the compound

eye (1), exoskeleton (2) and leg (3) evidence the enhanced image quality obtained using the propagation scheme for phase contrast imaging. (b) The figure depicts lineouts taken from the yellow shaded areas in figure 2b (solid black), c (dashed gray) and d (solid gray). Improved contrast and resolution at the few pixel level (≃70 μm) is achieved through edge enhancement in the phase contrast geometry (compare gray lines).

**Figure 3:** (a) A point source emits spherical x-ray wavefronts which are distorted on passing through a phase object. Propagating the distorted wavefront onto the detector, can lead to local focusing and defocusing. (b) Phase contrast image of a yellow jacket taken in a single shot 30 fs exposure of synchrotron radiation from the tabletop source, using the propagation technique of phase contrast imaging (a).

**Methods:**

**Wakefield Acceleration and Radiation Generation** The x-ray source discussed here is based on focusing a pulsed high power laser into a millimeter-sized plume of helium gas, which is immediately ionized and turned into a plasma. As the laser propagates through the plasma, it drives an electron density oscillation (plasma wave) with phase velocity near the speed of light in vacuum. The ponderomotive force of the laser displaces electrons from the almost stationary ions, setting up large accelerating fields. Electrons can be trapped by these fields, resulting in Gigaelectronvolt per centimeter energy gain[16-18]. At the same time, the electrons are oscillating transversely due to radial electrostatic fields of the plasma wave, emitting a bright beam of synchrotron-like x-rays[19] with appreciable degree of spatial coherence due to its micrometer source size[3].

**Laser** The experiments were carried with the high power HERCULES laser at the Center for Ultrafast Optical Science at the University of Michigan, Ann Arbor. A schematic is of the experimental setup is shown in figure 1a. Laser pulses with a pulse duration of $t_L$=32 fs and an energy of $E_L$=(2.2±0.1) J were focused to $d_{fwhm}$=(10.8±0.5) μm (full width at half maximum) onto the front edge of a supersonic

Helium gas jet with 3 mm diameter, reaching intensities of (2.0±0.4) Wcm$^{-2}$ and fully ionized plasma densities of 3 to 8×10$^{18}$ cm$^{-3}$. Quasi mono- and polyenergetic electron beams of ≃100 pC charge and peak energy of ≃120 MeV are deflected, and dispersed with a permanent magnet for measurement.

**Phase Contrast Imaging** In x-ray imaging, spatial contrast is a consequence of changes of the thickness and refractive index of the specimen $n=1-\delta-i\beta$, where $\delta$ and $\beta$ are the real and imaginary part of the refractive index. Changes of $\beta$ and $\delta$ integrated along the x-ray propagation direction and thickness of the specimen result in absorption and phase contrast respectively. Figure 3a indicates how phase contrast is achieved. The local propagation direction of an electromagnetic wave is perpendicular to the phase front (arrows). On passing through a phase object, spatial variations in $\delta$ will distort an initially spherical x-ray wavefront from a point source. Propagating the distorted wavefront a sufficient distance $v$ will lead to local focusing (converging arrows) of the x-rays, which can be observed as edge enhancement on an area detector. Monochromatic x-rays are not required[20]. To benefit from phase contrast enhancement, sufficiently spherical wavefronts, i.e. sufficient transverse coherence $L_{trans}$ is required. Transverse coherence $L_{trans} = \lambda u/(2\pi w)$ scales with the distance from the source $u$ and the inverse of the source size $w$, where $\lambda$ is the wavelength of the radiation. In our case, $L_{trans}$ is approximately 10 μm for 10 keV x-rays and the 1 μm source at only $u$=0.5 m from the source, allowing for phase contrast enhancement in a very compact geometry.

**Figure 1:**

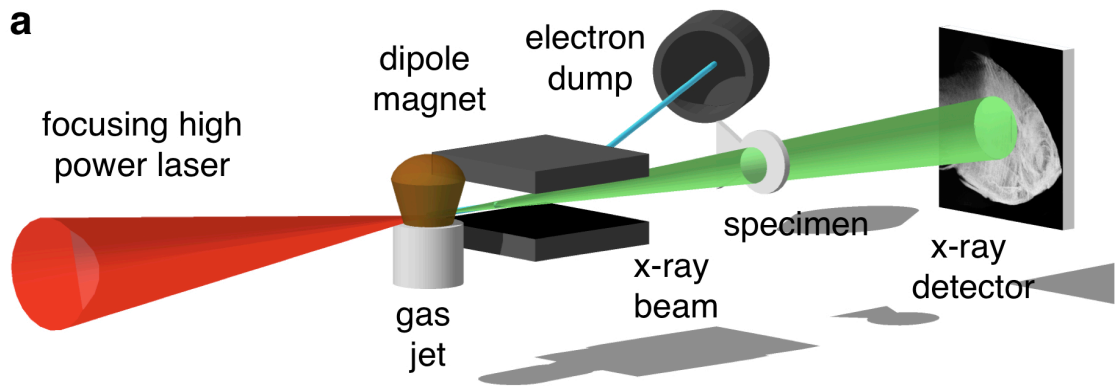

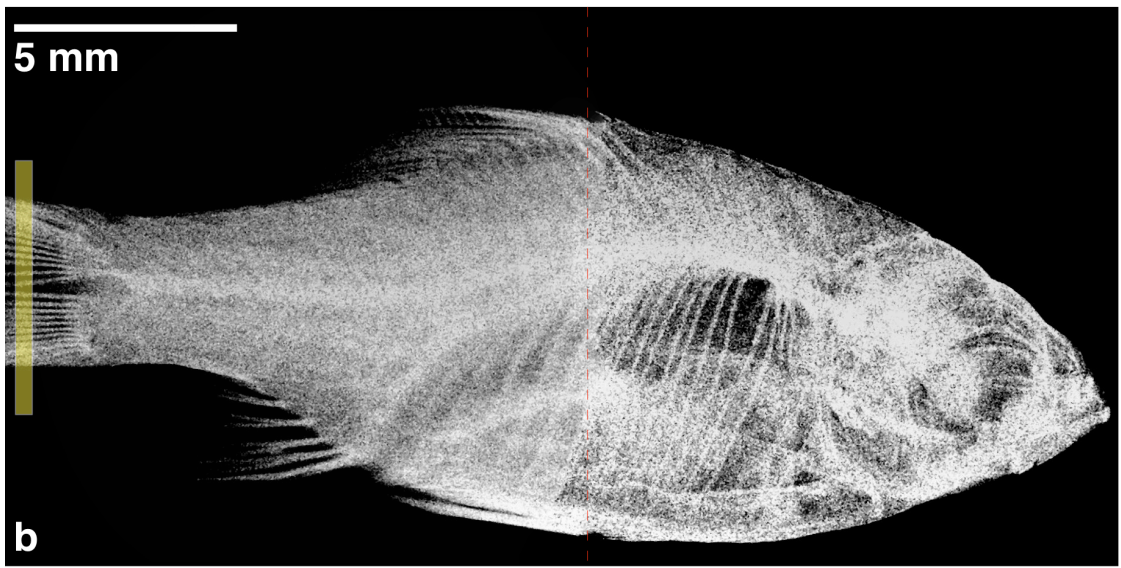

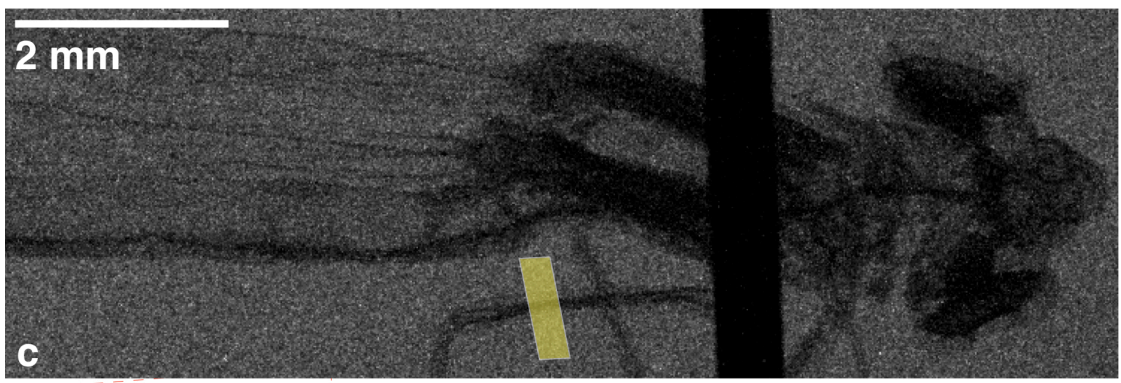

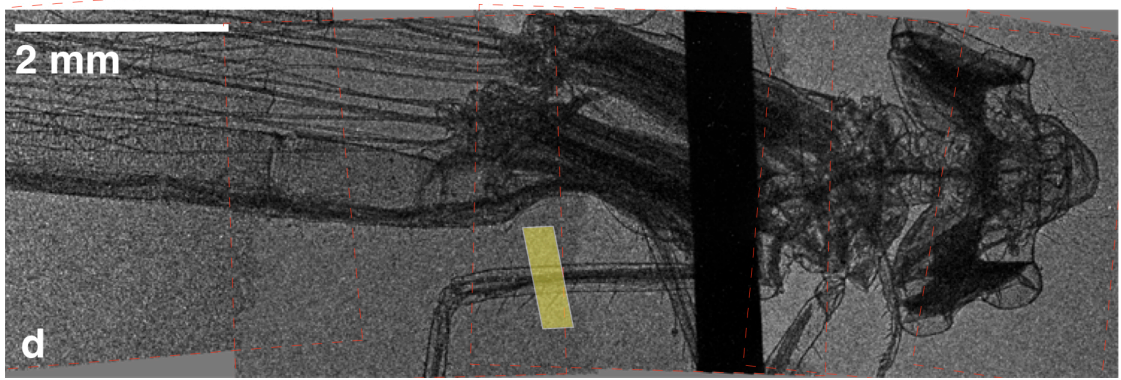

**Figure 2:**

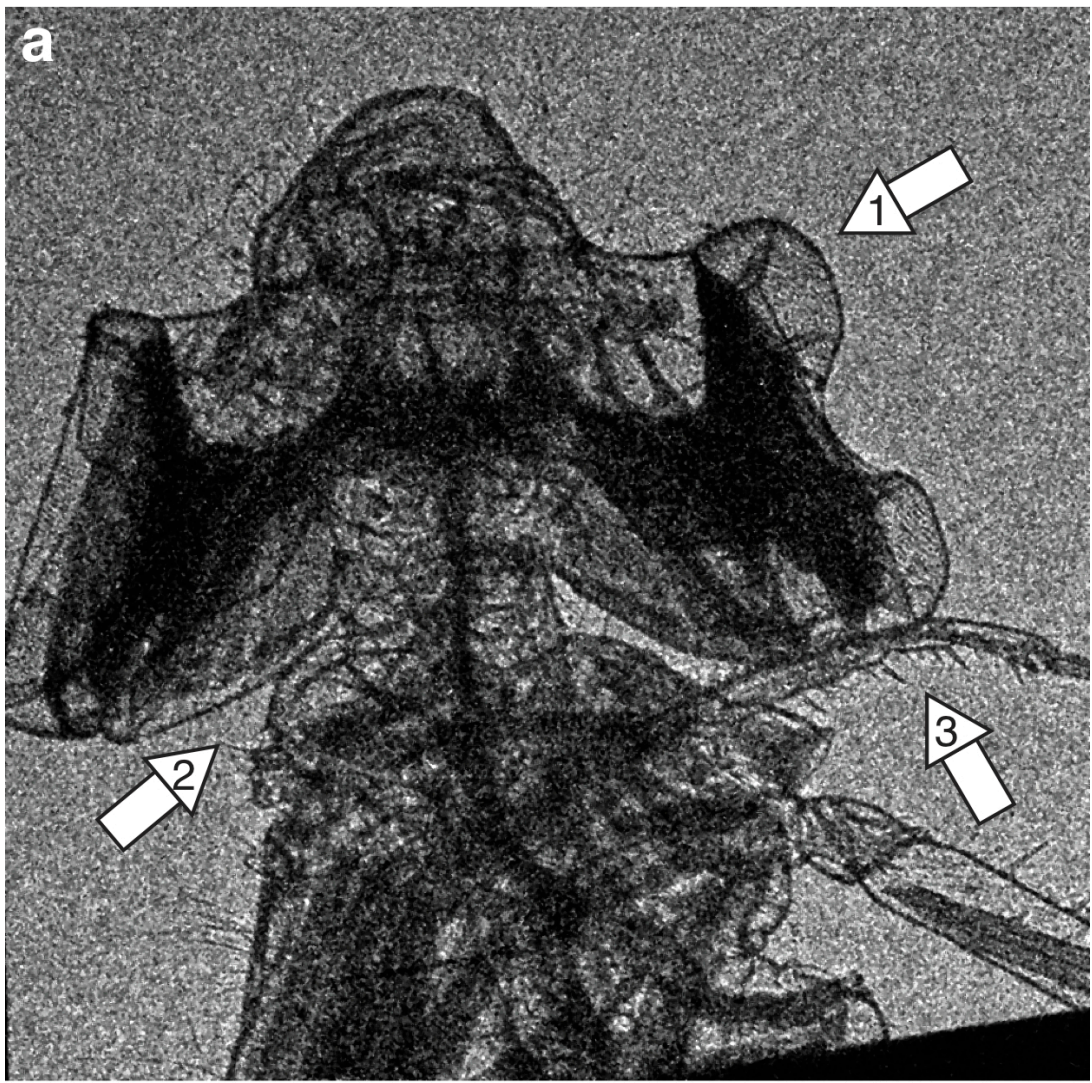

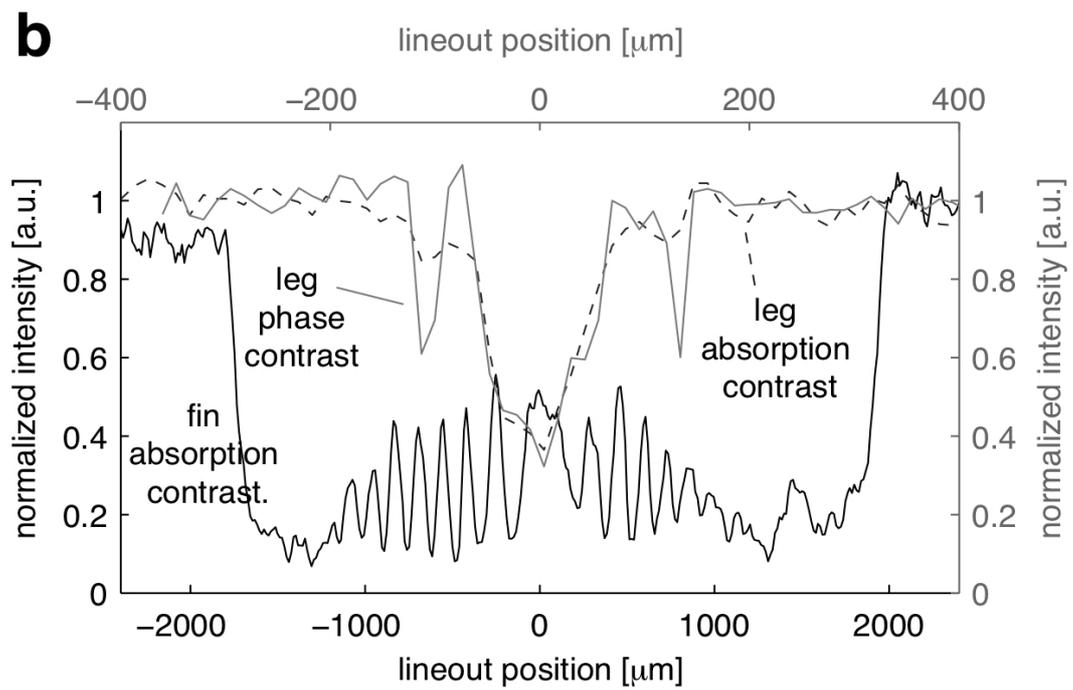

**Figure 3:**

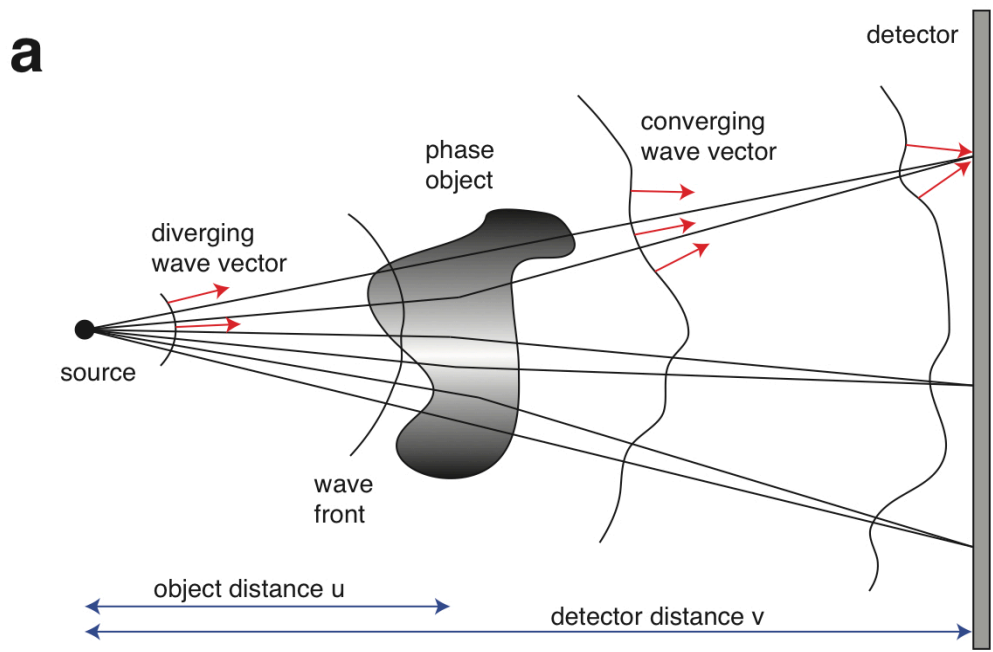

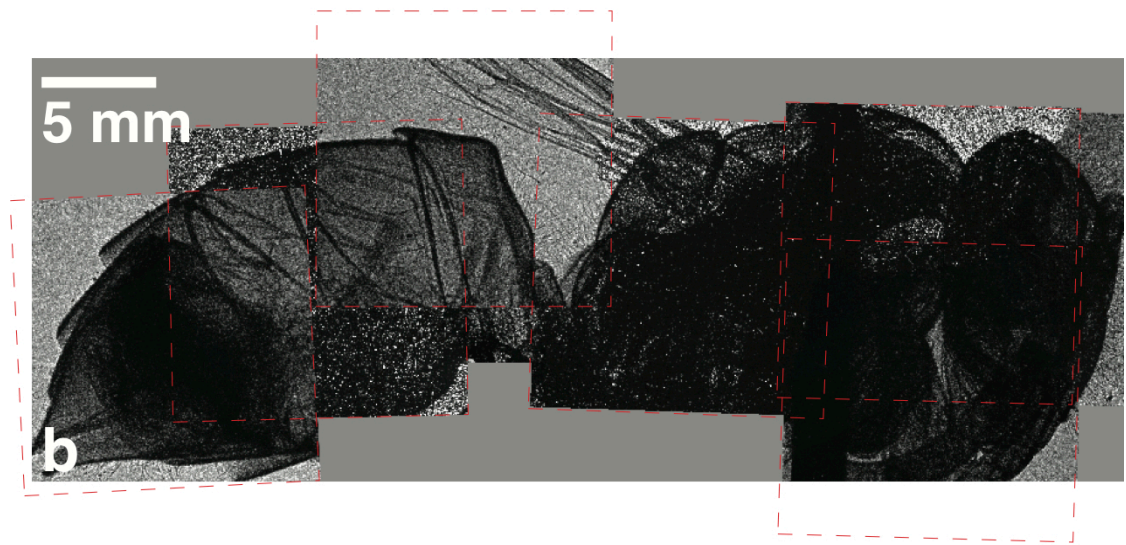